\documentstyle[aps,12pt]{revtex}
\newcommand{\clk}{C^\infty (\Lambda^k )}
\newcommand{\clpq}{C^\infty (\Lambda^{p,q})}
\newcommand{\cD}{{\cal D}}
\newcommand{\nhm}{{n-\frac m2}}
\begin{document}
\draft
\title{Heat kernel for non--minimal operators on a K\"{a}hler
manifold}
\author{Sergei Alexandrov}
\address{Department of Theoretical Physics, St.Petersburg University,
198904 St.Petersburg, Russia}
\author{Dmitri Vassilevich\thanks{On leave from:
Department of Theoretical Physics, St.Petersburg University,
198904 St.Petersburg, Russia. E.-mail: vasilevich@phim.niif.spb.su}}
\address{Insitut f\"{u}r Theoretische Physik, Technische Universit\"{a}t
Wien, Wiedner Hauptstr. 8-10, A-1040 Wien, Austria}
\date{Preprint TUW-96-01}
\maketitle
\begin{abstract}
The heat kernel expansion for a general non--minimal
operator on the spaces $\clk$ and $\clpq$ is studied. The
coefficients of the heat kernel asymptotics for this operator
are expressed
in terms of the Seeley coefficients for the Hodge--de Rham
Laplacian.
\end{abstract}

\section{Introduction}

Let $M$ be a compact Reimannian manifold of dimension $m$ without
boundary. If $M$ is equipped with integrable complex structure,
one can split tagential indices into holomorphic and antiholomorphic
ones and define space of differential forms $\clpq$. The
exterior differential $d$ can be also splitted in a sum
$d=\partial +\bar \partial$ of anticommuting nilpotent
operators: $\partial^2=\bar \partial^2=\partial \bar \partial +
\bar \partial \partial =0$. If $M$ is a K\"{a}hler manifold, the
corresponding ``Laplacians'' can be reduced to the Hodge--de Rham
Laplacian:
\begin{equation}
\partial \partial^*+\partial^*\partial =
\bar \partial \bar \partial^*+\bar \partial^*\bar \partial
=\frac 12 \Delta =\frac 12 (\delta d+d\delta ).
\nonumber
\end{equation}

Using these first order differential operators one can construct
a (non--minimal) second order differential operator:
\begin{equation}
\cD =g_1\partial \partial^*+g_2 \partial^*\partial +
g_3\bar \partial \bar \partial^* +g_4\bar \partial^*
\bar \partial +g_5\partial \bar \partial^*+
g_5^*\bar \partial \partial^* \label{eq:gener}
\end{equation}
with real constants $g_1,...,g_4$ and a complex constant
$g_5$. For some values of the constants this operator
reduces to that considered previously in the paper
\cite{GBF}, where one can find some motivations for
studying non--minimal operators. Such operators appear
naturally in quantum gauge theories after imposing
gauge conditions \cite{o1,o2,ChK,DNPB}.

For a self--adjoint second order operator $L$ with
non--negative eigenvalues $\{ \lambda_\nu \}$ one can
define the integrated heat kernel
\begin{equation}
{\rm Tr} (e^{-tL})=\sum_\nu e^{-t\lambda_\nu}.
\end{equation}
As $t\to 0+$, there is an asymptotic expansion of the
form
\begin{equation}
{\rm Tr} (e^{-tL})=\frac 1{(4\pi )^{\frac m2}}
\sum_{n=0}^\infty a_n(L)t^{\frac {(2n-m)}2} .
\end{equation}
In this paper we study the heat kernel expansion for the
non--minimal operator $\cD$ (\ref{eq:gener}) and relate
the Seeley coefficients $a_n(\cD )$ to that for the Laplace
operator $a_n(\Delta )$. General expressions for $a_n(\Delta )$,
$n=0,1,2,3$ suitable for the spaces of differential forms can be
found in the paper \cite{PBG}. In particular cases this
problem was solved in Refs. \cite{GBF,o2,ChK,DNPB}. In a sense,
we suggest an extension of the Theorem 1.2 of Ref. \cite{GBF}
for the case of complex geometry.

In the next section we study the heat kernel for $\cD$ acting
on the space of $k$-forms, $\clk$. In the section 3 the case
$\cD : \clpq \to \clpq$ is considered. This assumes some
restrictions on the constants in (\ref{eq:gener}), but a
more detailed information can be obtained. In the Appendix
we use complex projective space $CP^2$ as an example to check
up some of the equations of previous sections.

\section{Non--minimal operators on $k$-forms}

In this section we consider the heat kernel for non-minimal
operators acting on the space $\clk$ of $k$-forms.
First let us discuss some properties of first order operators
$D_1$ and $D_2$ which will be used later to build up a general
non--minimal second order operator $\cD$.

{\bf Lemma 1.} Let $D_1$ and $D_2$ be operators on
$C^\infty (\Lambda )$ having the following properties:
\newline
(a) $D_1, D_2 : C^\infty (\Lambda^k)\to C^\infty (\Lambda^{k+1})$,
(b) $D_1^2=D_2^2=0$, (c) $D_1D_2+D_2D_1=D_1D_2^*+D_2^*D_1=0$,
(d) $D_1D_1^*+D_1^*D_1=\alpha \Delta$,
$D_2D_2^*+D_2^*D_2=\beta \Delta$, $\alpha , \beta \ne 0$.
Then
\newline
1. $\clk ={\rm Ker} (\Delta )\oplus {\rm im}(D_1)\oplus {\rm
im}(D_1^*)
 ={\rm Ker} (\Delta )\oplus {\rm im}(D_2)\oplus {\rm
im}(D_2^*)$,
\newline
2. $\clk ={\rm Ker}(\Delta )\oplus (D_1D_2)_k\oplus (D_1D_2^*)_k
\oplus (D_1^*D_2)_k\oplus (D_1^*D_2^*)_k$.
\newline
3. The following mappings are isomorphisms:
\begin{eqnarray}
(D_1D_2)_k  {\stackrel{D_1D_1^*}{\leftrightarrow}} (D_1^*D_2)_{k-1},
&\quad &
(D_1^*D_2)_k  {\stackrel{D_2D_2^*}{\leftrightarrow}} (D_1^*D_2^*)_{k-1},
\nonumber \\
(D_1D_2^*)_k  {\stackrel{D_1D_1^*}{\leftrightarrow}} (D_1^*D_2^*)_{k-1},
&\quad &
(D_1D_2)_k  {\stackrel{D_2D_2^*}{\leftrightarrow}} (D_1D_2^*)_{k-1},
\nonumber
\end{eqnarray}
where operators $D$ act from right to left, and $D^*$ act from left to
right. We introduced the notation $(AB)_k={\rm im}(A)\cap {\rm im}(B)
\cap \clk$.

{\bf Proof}. The proof of the first statement repeats standard
proof \cite{stan} of the same property for operator $d$. The
decomposition 2. can be obtained by repeating twice the
decompositions 1. Last statement follows from the anticommutativity
properties b) and c).$\Box$

Let us introduce the following notations:
\begin{eqnarray}
\Delta_k=\Delta \vert_{\clk} , \quad f(t,D)={\rm Tr} \exp (-tD)
\nonumber \\
f(t,A,B,k)={\rm Tr}\exp (-t \Delta \vert_{(AB)_k}) , \nonumber \\
f_k(t)=f(t,D_1^*, D_2^*,k) \label{eq:2not}
\end{eqnarray}
$\beta_k$ denote Betti numbers.

{\bf Lemma 2.} $f_k(t)=\sum_{l=0}^k (-1)^l (l+1)
(f(t,\Delta_{k-l})-\beta_{k-l})$.

{\bf Proof}. First we observe that all spaces appearing in the second
statement of Lemma 1 are eigenspaces of the Laplace operator. Hence,
\begin{equation}
  f(t,\Delta_k )=\beta_k+f(t,D_1,D_2,k)+f(t,D_1^*,D_2,k)+
f(t,D_1,D_2^*,k)+f(t,D_1^*,D_2^*,k).
\label{eq:2.2}
\end{equation}
By using commutativity of $D_1$, $D_2$, $D_1^*$ and $D_2^*$ with
$\Delta$ and the last statement of Lemma 1 we obtain:
\begin{equation}
  \label{eq:2.3}
  f(t,D_1,D_2,k)=f(t,D_1^*,D_2,k-1)=f(t,D_1,D_2^*,k-1)=
f(t,D_1^*,D_2^*,k-2)=f_{k-2}(t)
\end{equation}
Now we can express $f_k(t)$ from (\ref{eq:2.2}) and by repeated
use of (\ref{eq:2.3}) obtain
\begin{equation}
  f_k(t)=-f_{k-1}+\sum_{l=0}^k (-1)^l (f(t,\Delta_{k-l})-
\beta_{k-l}).
\end{equation}
Now the statement of the Lemma follows by induction.$\Box$

It is easy to see that the operators
\begin{equation}
D_1=x_1\partial +y_1\bar \partial , \quad
D_2=x_2\partial +y_2\bar \partial \label{d1d2}
\end{equation}
satisfy conditions of Lemma 1 provided the equation
$x_1x_2^*+y_1y_2^*=0$ holds for complex parameters
$x_1$, $x_2$, $y_1$ and $y_2$. The constants $\alpha$
and $\beta$ are real and positive: $\alpha =\frac 12
(|x_1|^2+|y_1|^2)$, $\beta =\frac 12 (|x_2|^2+|y_2|^2)$.
The non-minimal operator
\begin{eqnarray}
\cD =aD_1D_1^*+bD_1^*D_1+cD_2D_2^*+dD_2^*D_2 \label{2.5} \\
=(a|x_1|^2+c|x_2|^2)\partial \partial^*+(b|x_1|^2+d|x_2|^2)
\partial^* \partial+(a|y_1|^2+c|y_2|^2)\bar \partial
\bar \partial^*+(b|y_1|^2+c|y_2|^2)\bar \partial^*
\bar \partial \nonumber \\
+((a-b)x_1y_1^*+(c-d)x_2y_2^*)\partial \bar \partial^*
+((a-b)y_1x_1^*+(c-d)y_2x_1^*)\bar \partial \partial^*
\nonumber
\end{eqnarray}
with real constants $a,b,c,d$ is the most general hermitian
operator on $\clk$ which can be constructed using
$\partial$, $\bar \partial$, $\partial^*$ and $\bar \partial^*$.
This operator has the form (\ref{eq:gener}).

The following Theorem represents the main result of this section.

{\bf Theorem 1.} Let $D_1$ and $D_2$ satisfy conditions of
Lemma 1. Then the coefficients $a_n$ of the heat kernel
expansion for the operator $\cD$ (\ref{2.5}) have the form:
\begin{eqnarray}
a_n(\cD \vert_{\clk})=(\alpha a+\beta c)^{\nhm}
\sum_{l=0}^{k-2} (-1)^{k-l}(k-l-1) a_n(\Delta_l)-
\nonumber \\
-((\alpha a+\beta d)^\nhm +(\alpha b+\beta c)^\nhm )
\sum_{l=0}^{k-1}(-1)^{k-l}(k-l)a_n(\Delta_l)+
\nonumber \\
+(\alpha  b+\beta d)^\nhm \sum_{l=0}^k (-1)^{k-l}
(k-l+1) a_n(\Delta_l ) . \nonumber
\end{eqnarray}

{\bf Proof.} By making use of Lemma 1 and equation (\ref{eq:2.3})
we find:
\begin{eqnarray}
f(t,\cD )=\beta_k +f(t,aD_1D_1^*+cD_2D_2^*,D_1,D_2,k)+
f(t,aD_1D_1^*+D_2^*D_2,D_1,D_2^*,k)+\nonumber \\
+f(t,bD_1^*D_1+cD_2D_2^*,D_1^*,D-2,k)+
f(t,bD_1^*D_1+dD_2^*D_2,D_1^*,D_2^*,k) \nonumber \\
=\beta_k+f(t,(b\alpha +c \beta )\Delta, D_1,D_2,k)+
f(t,(a\alpha +d\beta )\Delta ,D_1,D_2^*,k)+\nonumber \\
+f(t,(b\alpha +c\beta )\Delta ,D_1^*,D_2,k)+
f(t,(b\alpha +d\beta )\Delta ,D_1^*,D_2^*,k)\nonumber \\
=\beta_k +f_{k-2}((a\alpha +c\beta )t)+
f_{k-1}((a\alpha +d\beta )t)+f_{k-1}((b\alpha +c\beta )t)+
f_k(b\alpha +d\beta )t) \nonumber
\end{eqnarray}
Now we can use Lemma 2 in order to express the last line
in terms of the Laplace operators on $\clk$.
\begin{eqnarray}
f(t,\cD )=\beta_k +\sum_{l=0}^k (-1)^l (l+1)
(f((b\alpha +d\beta )t,\Delta_{k-l})-\beta_{k-l}+
\nonumber \\
+\sum_{l=0}^{k-1}(-1)^l[f((a\alpha +d\beta )t,\Delta_{k-l-1})
+f((b\alpha +c\beta )t,\Delta_{k-l-1})-2\beta_{k-l-1}]+
\nonumber \\
+\sum_{l=0}^{k-2}(-l)^l(l+1)(f((a\alpha +c\beta )t,\Delta_{k-l-2})
-\beta_{k-l-2}) . \nonumber
\end{eqnarray}
One can see that the contributions of the Betti numbers
to different sums cancel each other. By making asymptotic
expansion of the last equation we arrive at the statement
of this Theorem. $\Box$

Note, that to ensure existence of all traces of exponentials
for positive $t$ one should take non-negative $a,b,c,d$.

\section{Non-minimal operators on $(p,q)$-forms}

This section is devoted to non--minimal operators on
$\clpq$. To ensure that $\cD$ maps $\clpq$ on itself
we should choose
\begin{equation}
D_1=\partial ,\quad D_2=\bar \partial .
\end{equation}
The following notations will be useful:
\begin{eqnarray}
(AB)_{p,q}={\rm im}(A)\cap {\rm im}(B)\cap \clpq ,
\nonumber \\
\Delta_{p,q}=\Delta \vert_{\clpq}.
\label{notpq}
\end{eqnarray}
Other notations are modified by replacing $k$ by $p,q$
in (\ref{eq:2not}). $\beta_{p,q}$ will denote Hodge numbers.

Next Lemma replaces the Lemma 1.

{\bf Lemma 3.} 1. $\clpq ={\rm Ker}(\Delta_{p,q})
\oplus (\partial \bar \partial )_{p,q} \oplus
(\partial \bar \partial^*)_{p,q}\oplus
(\partial^* \bar \partial )_{p,q}\oplus
(\partial^* \bar \partial^*)_{p,q}$.
\newline
2. The following maps are isomorphisms:
\begin{eqnarray}
(\partial \bar \partial )_{p,q}
\stackrel{\partial ,\ \partial^*}{\leftrightarrow}
(\partial^*\bar \partial )_{p-1,q} &\qquad&
(\partial^*\bar \partial )_{p,q}
\stackrel{\bar \partial ,\ \bar \partial^*}{\leftrightarrow}
(\partial^* \bar \partial^* )_{p,q-1} \nonumber \\
(\partial \bar \partial^*)_{p,q}
\stackrel{\partial ,\ \partial^*}{\leftrightarrow}
(\partial^* \bar \partial^*)_{p-1,q} &\qquad&
(\partial \bar \partial )_{p,q}
\stackrel{\bar \partial ,\ \bar \partial^*}{\leftrightarrow}
(\partial \bar \partial^*)_{p,q-1} \nonumber
\end{eqnarray}
where the operators $\partial$ and $\bar \partial$ act from
right to left, and the operators $\partial^*$ and
$\bar \partial^*$ act from left to right.

The proof repeats that of Lemma 1.

{\bf Lemma 4}. $f_{p,q}(t)=\sum_{k,l=0}^{p,q} (-1)^{k+l}
(f(t,\Delta_{p-k,q-l})-\beta_{p-k,q-l})$.

{\bf Proof.} Lemma 3 gives the following identities:
\begin{eqnarray}
f(t,\Delta_{p,q})=\beta_{p,q}+f(t,\partial ,\bar \partial ,p,q)
+f(t,\partial ,\bar \partial^*,p,q)+
f(t,\partial^*,\bar \partial ,p,q)+
f(t,\partial^*,\bar \partial^*,p,q), \nonumber \\
f(t,\partial ,\bar \partial ,p,q)=
f(t,\partial^*,\bar \partial ,p-1,q)=
f(t,\partial ,\bar \partial^*,p,q-1)=
f(t,\partial^*,\bar \partial^*,p-1,q-1)
\nonumber
\end{eqnarray}
Thus we obtain:
\begin{equation}
f_{p,q}(t)=f(t,\Delta_{p,q})-\beta_{p,q}-f_{p-1,q}(t)
-f_{p,q-1}(t)-f_{p-1,q-1}(t). \label{eq:L4}
\end{equation}
Repeated use of (\ref{eq:L4}) gives the statement of Lemma 4.$\Box$

Now one can prove the following Theorem.

{\bf Theorem 2}. Let $\cD =a\partial \partial^*+b\partial^*\partial
+c\bar \partial \bar \partial^*+d\bar \partial^*\bar \partial$ on
$\clpq$. Then
\begin{eqnarray}
a_n(\cD )=\left (\frac {b+d}2 \right )^\nhm a_n(\Delta_{p,q})
+\left ( \left ( \frac {b+c}2 \right )^\nhm -
\left ( \frac {b+d}2 \right )^\nhm \right ) a_n(\Delta_{p,q-1})
\nonumber \\
+\left ( \left ( \frac {a+d}2 \right )^\nhm -
\left ( \frac {b+d}2 \right )^\nhm \right ) a_n(\Delta_{p-1,q})
\nonumber \\
+\left ( \left ( \frac {a+c}2 \right )^\nhm +
\left ( \frac {b+d}2 \right )^\nhm -
\left ( \frac {a+d}2 \right )^\nhm -
\left ( \frac {b+c}2 \right )^\nhm \right )
\sum_{k,l=0}^{p-1,q-1} (-1)^{p+q-k-l}a_n(\Delta_{k,l}).
\nonumber
\end{eqnarray}

As in the previous section we expressed the Seeley coefficients
for a non--minimal operator in terms of the Seeley coefficients
of the Laplacian.

\section{conclusions}

In this paper we expressed the heat kernel coefficients for
non--minimal operators on $\clk$ and $\clpq$ in terms of the
Seeley coefficients for the Laplace operators on the same
spaces. Expressions for the heat kernel asymptotics applicable
for Laplacian on differential forms can be found in the
literature \cite{PBG}.

The fact that underlying manifold is K\"{a}hlerian was
used only to relate $\partial \partial^*+\partial^*\partial$
and $\bar \partial \bar \partial^*+\bar \partial^*\bar \partial$
to $\Delta =\delta d+d\delta$. With some modifications our
results can be extended to a general complex manifold.
Another generalization could consist in adding an endomorphism
$E$ to the operator $\cD$.

\section*{Acknowledgments}

One of the authors (DV) is grateful to W. Kummer for warm
hospitality at the Technische Universit\"{a}t Wien. This work
was partially supported by Fonds zur F\"{o}rderung der
wissenschaftlichen Forschung, project P10221-PHY.

\section*{Appendix: Consistency check. $CP^2$}

First let us construct the harmonic expansion for $(p,q)$ forms
on $CP^2=SU(3)/SU(2)\times U(1)$. All needed material on structure and
geometry of $CP^2$, as well as a method of constructing of harmonic
expansion, can extracted from ref. \cite{CP2}.

For any homogeneous space $G/H$ a field $\Phi$ belonging to
an irreducible representation $D(H)$ of $H$ can be decomposed
in a sum of harmonics corresponding to all representations of
$G$ giving $D(H)$ after reduction to $H$. In the case of $CP^2$
the representations $D(SU(2)\times U(1))$ corresponding to
$(p,q)$ forms can be easily found:
\begin{eqnarray}
(p,q)=(0,0) &\quad & D(SU(2)\times U(1))=(1,0) \nonumber \\
(p,q)=(0,1) &\quad & D(SU(2)\times U(1))=(2,-1) \nonumber \\
(p,q)=(1,0) &\quad & D(SU(2)\times U(1))=(2,1) \nonumber \\
(p,q)=(0,2) &\quad & D(SU(2)\times U(1))=(1,-2) \nonumber \\
(p,q)=(2,0) &\quad & D(SU(2)\times U(1))=(1,2) \nonumber \\
(p,q)=(1,1) &\quad & D(SU(2)\times U(1))=(1,0)\oplus (3,0)
\label{eq:doth}
\end{eqnarray}
The representations of $SU(2)\times U(1)$ are labelled by
dimension of $SU(2)$ representation (first number) and the
$U(1)$ charge (second number). The representations for other
values of $p$ and $q$ can be restored by using the duality
transformation. Note that the representation in last line
of (\ref{eq:doth}) is reducible.

By repeating calculations of Ref. \cite{CP2} one can find
the $SU(3)$ representations contributing to the harmonic
expansion:
\begin{eqnarray}
(p,q)=(0,0) &\quad & D(SU(3))=(m,m),\ m=0,1,\dots \nonumber \\
(p,q)=(0,1) &\quad & D(SU(3))=(m,m)\oplus (n,n+3), n=0,1,\dots
\ m=1,2\dots \nonumber \\
(p,q)=(1,0) &\quad & D(SU(3))=(m,m)\oplus (n+3,n), n=0,1,\dots
\ m=1,2,\dots \nonumber \\
(p,q)=(0,2) &\quad & D(SU(3))=(n,n+3),\ n=0,1,\dots \nonumber \\
(p,q)=(2,0) &\quad & D(SU(3))=(n+3,n),\ n=0,1,\dots \nonumber \\
(p,q)=(1,1) &\quad & D(SU(3))=(n,n)_1\oplus (n,n)_2\oplus (0,0)
\oplus (m,m+3) \oplus (m+3,m) \nonumber \\
\ m=0,1,\dots , \quad n=1,2,\dots \label{eq:dotg}
\end{eqnarray}
where the representations of $SU(3)$ are labelled by their
Dynkin indices. In the last line subscripts are introduced
in order to distinguish between equivalent representations.

Let $V^{p,q}(m,n)$ denotes the space of $(p,q)$-forms transforming
according to the representation $(m,n)$ of $SU(3)$. One can easily
prove that
\begin{eqnarray}
\partial V^{0,0}(m,m)=V^{1,0}(m,m), \quad
\bar \partial V^{0,0}(m,m)=V^ {0,1}(m,m), \quad
m\ge 1 ,\nonumber \\
\partial V^{0,0}(0,0)=\bar \partial V^{0,0}(0,0)=\{ 0\} ,
\nonumber \\
\bar \partial V^{1,0}(m,m)=\partial V^{0,1} (m,m)=
V^{1,1}(m,m)_1, \nonumber \\
\partial V^{1,0}(m,m)=\bar \partial V^{0,1}(m,m)=
\{ 0\} , \nonumber \\
\partial V^{1,0}(n+3,n)=V^{2,0}(n+3,n), \quad
\bar \partial V^{0,1}(n,n+3)=V^{0,2}(n,n+3),
\nonumber \\
\bar \partial V^{1,0}(n+3,n)=V^{1,1}(n+3,n), \quad
\partial V^{0,1}(n,n+3)=V^{1,1}(n,n+3),
\nonumber
\end{eqnarray}
and so on. Eigenvalues of the Laplace operator coincide with
values of the quadratic Casimir operator $C_2$ of $SU(3)$ in
corresponding representation. Degeneracy of an eigenvalue
$C_2(m,n)$ is given by dimension $d(m,n)$ of the representation
$(m,n)$. For the sake of completeness, we give here explicit
expressions for $C_2$ and $d$, though they will not be used
in what follows.
\begin{eqnarray}
C_2(m,n)=\frac 13 (m^2+n^2+mn+3m+3n)\nonumber \\
d(m,n)=\frac 12 (m+1)(n+1)(m+n+2) \nonumber
\end{eqnarray}
The heat kernels become
\begin{eqnarray}
f(t,\Delta_{00})&=& \sum_{m=0}^\infty d(m,m) \exp (-tC_2(m,m)),
\nonumber \\
f(t,\Delta_{0,1})&=& \sum_{m=1}^\infty d(m,m) \exp (-tC_2(m,m))+
\nonumber \\
 &+& \sum_{n=0}^\infty d(n,n+3) \exp (-tC_2(n,n+3)),\ \ {\rm etc}
\nonumber
\end{eqnarray}
With these formulae at hand one can check up Lemma 2 and Lemma 4.


%
%

\end{document}